\documentclass[10pt,a4paper,twocolumn]{article}
\usepackage[dvips]{graphicx}
\usepackage{subfigure}
\usepackage{amssymb, amsmath}
\usepackage{citesupernumber}
\usepackage{citecollapse}
\usepackage[small,bf]{caption}
\usepackage{setspace}
\usepackage{graphicx}

\title{A three-species model explaining cyclic dominance of pacific salmon}
\author{C. Guill\thanks{Institute of Condensed Matter Physics, Darmstadt University of Technology, Hochschulstr.~6, 64289 Darmstadt, Germany}, 
B.~Drossel$^*$, W. Just\thanks{School of Mathematical Sciences, Queen Mary University of London, London E1 4NS, UK}, E. Carmack\thanks{Institute of Ocean Sciences, 9860 West Saanich Road
Sidney B.C. V8L 4B2 Canada}}

\begin{document}
\maketitle

\begin{abstract}The four-year oscillations of the number of
spawning sockeye salmon (\textit{Oncorhynchus nerka}) that return to their native stream within
the Fraser River basin in Canada are a striking
example of population oscillations. The
period of the oscillation corresponds to the dominant generation
time of these fish. Various - not fully convincing -
explanations for these oscillations have been proposed,
including stochastic influences, depensatory fishing, or genetic effects.
Here, we show that the oscillations can be explained as a
stable dynamical attractor of the population dynamics, resulting
from a strong resonance near a Neimark Sacker bifurcation. This
explains not only the long-term persistence of these oscillations,
but also reproduces correctly the empirical sequence of salmon abundance within one period of the oscillations.
Furthermore, it explains the observation that these oscillations occur only in sockeye stocks originating from large oligotrophic lakes, and that they are usually not observed in salmon species that have a longer generation time.\end{abstract}


\section{Introduction}
In many ecological systems distinct population oscillations are known, with their specific dynamical characteristics often captured by simple generic models. Among these are the spatial synchronisation of the lynx-hare oscillations in Canada \cite{lynx1,lynx2,lynx3}, the chaotic oscillations of boreal rodents in Fennoscandia \cite{rodent1}, or the cyclic outbreak dynamics of the spruce budworm \cite{spruce1,spruce2}.
The four-year oscillations of sockeye salmon (\textit{Oncorhynchus nerka}) in the Fraser River basin in Canada are another well-documented example of such large-scale population oscillations \cite{ricker1950,townsend,ricker1997}. Every fourth year the abundance of these fish is at very high levels, reaching several million fish in some spawning populations, but drops to numbers between several hundred and a few ten thousand individuals in the following years (hence the term \textit{cyclic dominance}). Different stocks can have their population maximum in different years (Fig. 1 and Fig. A1 in Appendix A). The oscillations were reported as early as the
19th century and are evident for instance in the extremely high
catches by fisheries every fourth year\cite{rounsefell}. This both economically and conservationally important phenomenon has been ascribed either to transient effects or to stochastic influences\cite{myers}, to depensatory predation\cite{larkin}, to fishing\cite{walters-staley}, or to genetic effects\cite{levy,walters-woodey}, but all of these explanations are still not fully convincing \cite{levy,ricker1997}.

The sockeye salmon return to spawn in their
native stream or lake in late summer and then die, which means that the salmon generations do not overlap. 
The hatched fry migrate downstream in the following spring and feed for one season in large freshwater lakes, before they migrate to the ocean, where they spend the next two and a half years.
The carcasses of the adult salmon are decomposed and provide an important phosphorus input into the rearing lakes of the fry. Since the oscillations of sockeye salmon originating from different lakes are not in phase, we have clear evidence that the relevant processes causing the phenomenon of cyclic dominance take place in the rearing lakes rather than in the ocean.

\begin{figure}[h!t!b!p]
\begin{center}
 \includegraphics[scale=1]{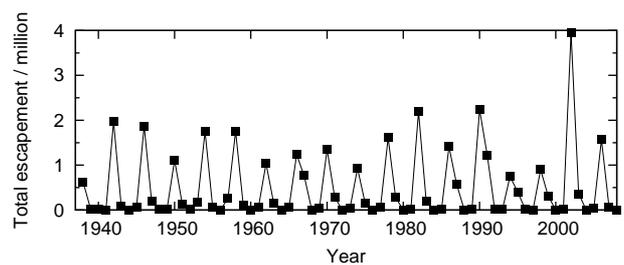}
\caption{\textbf{Time series of sockeye salmon abundance.} Total escapement (number of male and female adults that escape the fisheries) of the late Shuswap Lake run. Shuswap lake is one of the most important rearing lakes of juvenile sockeye salmon. Data for 5 other lakes are provided in Appendix A.}
\label{fig:empirical_data}
\end{center}
\end{figure}

\section{Model and results of computer simulations}
It is our aim to develop a generic model capturing only the essential mechanisms required for the occurrence of cyclic dominance. As such, the model is kept very simple and is not designed to quantitatively predict the population dynamics of all species in the corresponding lake ecosystems.

The model uses standard continuous population dynamics equations for
sockeye fry, $s_n(t)$, their predator (e.g. rainbow trout), $p_n(t)$, and their zooplankton
food (mainly daphnia), $z_n(t)$, during the growth season from spring ($t=0$) to fall ($t=T$) in year $n$,
combined with a rule for calculating the three population sizes at the
beginning of the next season as a function of the population sizes at
the end of the previous season(s) (see \textit{Materials and Methods}). The sockeye fry population at the
end of a season gives rise to the number of spawning adults 3 years
later, which in turn determines the number of sockeye fry in the
following spring, and represents a nutrient input that affects the
carrying capacity of the zooplankton. A small fraction of the salmon
stays in the ocean for one more year and mature at age 5, thus
causing a mixing between the four brood lines of a spawning
population. An even smaller fraction matures at age 3, but
since these fish are predominantly small-sized males\cite{ricker1997}
that do not influence the number of fertilised eggs, they are
neglected in this study.

According to empirical observations\cite{levy}, the zooplankton level at the end of one year has no effect on the following year. However, its carrying capacity is a function of the nutrients provided by the adult salmon (and thus a function of the initial fry biomass of that year): $K_n=K_0+f(s_n(0))$, where $K_0$ is the carrying capacity in the absence of sockeye spawners, and $f$ is a function that increases with the number of spawners. The zooplankton is initialised each year with its carrying capacity. The predator biomass does not change over the winter season in our model.

Fig. 2a shows a time series of the biomass of the sockeye fry at the end of the growth season. The first 300 years are cut off to show only the stable periodic
oscillation with one strong year followed by one intermediate year and two weak years, just as
in the empirical data of most sockeye stocks showing cyclic dominance (Fig. 1 and Appendix A). When the simulation parameters are chosen differently, the difference between the strong and the intermediate year may become less pronounced, or the order may even become reversed. This is also observed in the empirical data of some stocks.
\begin{figure}
\begin{center}
\includegraphics[scale=1]{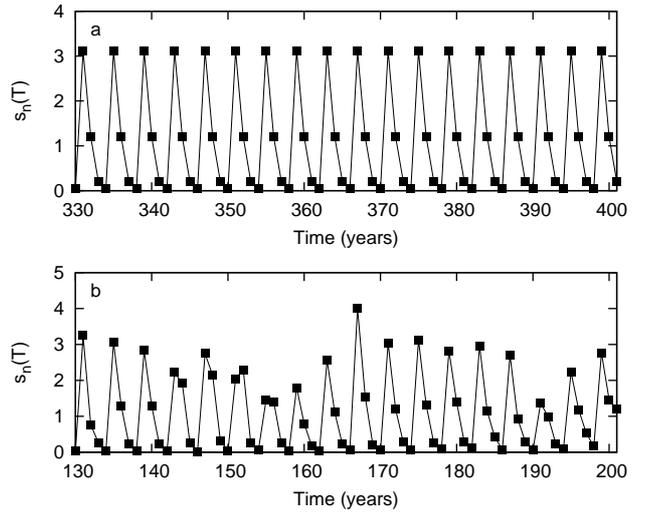}
\caption{\textbf{Simulated time series}. Stable attractor of sockeye salmon dynamics generated by computation of the time-continuous model, showing oscillations with period 4 and the same pattern in the sequence of salmon abundance as in the empirical data. Only the biomasses at the end of each growth season are shown. a: deterministic model, b: dynamics with up to 50 percent fluctuations in the survivability of the sockeye salmon in the ocean (see \textit{Discussion}). Parameter values for these simulations are as explained in \textit{Materials and Methods}.}
\label{fig:timeseries}
\end{center}
\end{figure}

When a parameter is varied, for instance the constant fraction of the zooplankton carrying capacity, $K_0$, the dynamical pattern may change. Fig. 3 shows the biomass $s_n(T)$ of the sockeye fry at the end of the season for different values of $K_0$, from
year $t=$1000 to year 1100. For small $K_0$, all sockeye lines are
equally strong. This means that the dynamics reaches a fixed point and that there is no cyclic dominance. With increasing $K_0$, the fixed point eventually becomes unstable. First a bifurcation to quasiperiodic behaviour occurs, and then the frequency of the oscillation becomes locked at 4.

\begin{figure}
\begin{center}
 \includegraphics[scale=1]{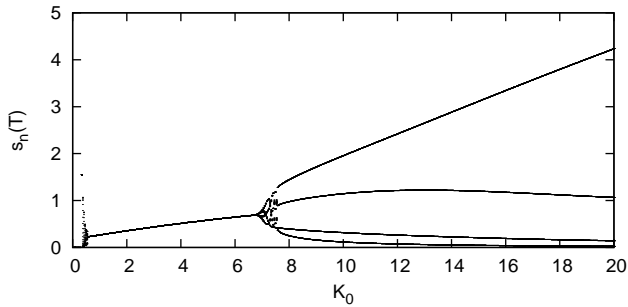}
\caption{\textbf{Bifurcation diagram of the biomass of the sockeye fry at the end of the season.} Bifurcation parameter is the constant fraction of the zooplankton carrying capacity $K_0$. For each value of $K_0$, 100 consecutive data points are plotted. Whenever only a small number of different points is visible for a value of the bifurcation parameter, this indicates periodic oscillations or a fixed point.}
\label{fig:bifurcation}
\end{center}
\end{figure}

\section{Linearised Theory}
In order to understand and interpret these observations, we first note
that the continuous population dynamics during the season, together
with the matching conditions applied between two seasons, can be
viewed as a discrete map (a so-called Poincar\'e map), giving the
sockeye and predator biomasses at the end of one year as function of
the biomasses at the end of the previous years.

To obtain this map,
one first has to integrate the dynamical equations over one season,
giving $s_n(T)$ and $p_n(T)$ as a function of $s_n(0)$ and $p_n(0)$.
The zooplankton can be eliminated since the initial condition of the
zooplankton depends on $s_n(0)$ only. Next, one expresses $s_{n+1}(0)$
and $p_{n+1}(0)$ as a function of $s_{n-3}(T)$, $s_{n-4}(T)$ and
$p_{n}(T)$, using the matching conditions.

The mechanism which generates the population oscillation is based on the
nature of the instability of the stationary state
$s_n(T)=s^*(T)$, $p_n(T)=p^*(T)$ of our system.
The corresponding bifurcation can be investigated in terms of
a linear stability analysis. Close to the fixed
point, the dynamics can be approximated by linear terms. Denoting
the distance of the biomasses from their fixed point value by $\delta
s_n=s_n(T)-s^*(T)$ etc, the linear approximation of this map has the form
\begin{equation}
\left(\begin{array}{c}\delta s_{n+1} \\ \delta s_{n} \\\delta s_{n-1}
\\\delta s_{n-2}\\\delta s_{n-3}\\\delta p_{n+1}
\end{array}\right)
=
\left(\begin{array}{cccccc}
0&0&0&m_s&\epsilon_1&-a\\
1&0&0&0&0&0\\
0&1&0&0&0&0\\
0&0&1&0&0&0\\
0&0&0&1&0&0\\
0&0&0&b&c&m_p
\end{array}\right)
\left(\begin{array}{c}\delta s_{n} \\ \delta s_{n-1} \\\delta s_{n-2}
\\\delta s_{n-3}\\\delta s_{n-4}\\\delta p_{n}
\end{array}\right)
\end{equation}
with positive parameters $m_s,m_p,a,b,c,\epsilon_1$. The general structure of the matrix in Eq. 1 is determined only by the matching conditions and is valid for any time-continuous model applied during the seasons. The equations of motion of the latter only determine the numerical values of the parameters in Eq. (1).

The first line of
the matrix describes the influence on the sockeye fry of year $n+1$ of
the sockeye fry of year $(n-3)$ and $(n-4)$ (which are the parents of the
fry in year $(n+1)$), and of the predator population. $m_s$ and
$\epsilon_1$ are positive, since more parents imply more
offspring. $-a$ is negative, since more predators imply less fry. The
other nontrivial line of this matrix, the last line, describes the
influence on the predators of year $(n+1)$ of the sockeye fry of year
$(n-3)$ and $(n-4)$ (which are the parents of the predator's food), and of
the predator population in the previous year. All three parameters are
positive, since more food implies more predator growth and since more predators in one year give rise to more predators in the next year.

The eigenvalues of this matrix determine the nature of the dynamics near the bifurcation. When all eigenvalues have an absolute value smaller than 1, the fixed point is stable, and the dynamics converges to this fixed point. When the absolute value of one or more eigenvalues is larger than 1, the fixed point is unstable, and the dynamics approaches a different attractor. 

In order to understand the dynamics resulting from this matrix, we
first consider the case that the matrix element $\epsilon_1$ and the
product $a(b+c)$ vanish. (Note that since all parameters are positive,
the latter requires both products $ab$ and $ac$ to vanish.) This means
that all salmon return at age 4, and that the trout have a good choice
of other food so that their dynamics is independent of that of the
salmon fry. In this case the eigenvalues of the matrix are $m_p$, 0,
and the four fourth roots of $m_s$. Since the four salmon lines are
independent from each other in this case, the sequence $\delta s_t$
has trivially the period 4 and simply iterates the initial four
values, with an amplitude decreasing for $m_s<1$ and increasing
otherwise (and with the trout being independent of the salmon).  When
$m_s$ is increased from a value smaller than 1 to a value larger than
1, all four eigenvalues $m_s^{1/4}$ cross the unit circle
simultaneously, and the fixed point becomes unstable.  This degeneracy
is lifted when the parameters $\epsilon_1$ and $a(b+c)$ are made
nonzero. As long as these parameters are not large, one can expect the
four main eigenvalues to remain close to the real and imaginary axis,
respectively, implying a (possibly damped) oscillation with a period
close to 4.

\begin{figure}
\begin{center}
 \includegraphics[scale=1, angle=0]{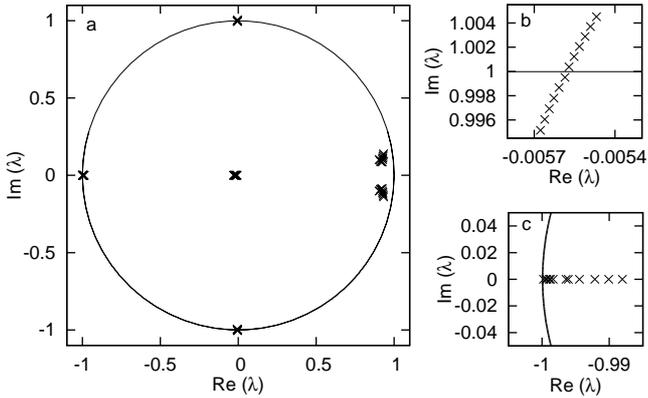}
 \caption{\textbf{Eigenvalues of the matrix from Eq. 4, as obtained from a
   linear stability analysis of the time-continuous model.} $K_0$ increases
   from 6.4 to 7.5 as the eigenvalues move outwards. The zoom shows that the eigenvalues close to the imaginary axis are the first ones to cross the unit cycle.}
\label{fig:eigenvalues}
\end{center}
\end{figure}

The type of bifurcation that occurs when the fixed point
becomes unstable depends on which eigenvalue crosses first the unit
circle as a control parameter is increased. In order to determine the
type of the bifurcation, we evaluated the parameters of the matrix
numerically from our computer simulations of the time continuous model
near the bifurcation, and we calculated the eigenvalues $\lambda$ of
the matrix. Figure \ref{fig:eigenvalues} shows that the bifurcation is
caused by a pair of complex conjugate eigenvalues crossing the unit
circle, indicating a Neimark Sacker bifurcation (the discrete version
of the Hopf bifurcation). Since the dominant eigenvalues are close to
$\pm i$, the period of the resulting oscillation is close to 4.  The
bifurcation is mainly driven by the parameters $m_s$ and, to a minor
extent, $a$, which correspond to reproduction of the 4-year old
sockeyes and predation by the rainbow trouts, respectively. The fixed
point becomes unstable when either of these parameters increases.

Now, it is known from the theory of bifurcations that if the period at
such a bifurcation is close to 4, there occurs a strong resonance,
which means that the period becomes locked exactly at the value 4 not
far beyond the bifurcation point. In contrast to conventional
frequency locking, a strong resonance is due to nonlinearities that
are of the same order as the leading nonlinearity, and frequency
locking therefore occurs over a much wider range of parameters\cite{kuznetsov}.

The two complex conjugate eigenvalues are the first ones to cross the
unit circle when $a$ and $\epsilon_1$ are small non-negative numbers
and when $m_p$ is well below 1. Since $\epsilon_1$ and $a(b+c)$ are
small compared to $m_s$, the period of the resulting oscillation
remains close to 4, and the locking at period 4 therefore occurs not
far beyond the bifurcation.  When these conditions are not met, the
unit circle is typically first crossed at -1, and a period-doubling
(or \textit{flip}) bifurcation occurs. 

\section{Discussion}
All these considerations lead to two basic conditions under which strong
resonance can be observed: First, an increase in the number of
spawning salmon must lead to a sufficiently strong increase in this
number four years later (i.e. $m_s$ must be large enough). If the
number of fry migrating to the ocean is dominated by other factors,
such as a strongly limited carrying capacity for the fry, the
population will be at a fixed point rather than on the oscillating
side of the bifurcation. Second, the four salmon lines must be coupled
in order for the Neimark Sacker bifurcation to occur, rather than a
period doubling bifurcation. In our model, this
coupling is due to a fraction of sockeye returning at age 5 instead of
age 4, and, more importantly, due to the predator being sufficiently
strongly coupled to sockeye dynamics. 

Previous studies of salmon dynamics, used in salmon management, are based on the Ricker model\cite{ricker1950,hume1996} or the Larkin
model\cite{larkin,martell2008,marsden2009}. Because both models have a
strongly limited carrying capacity for the fry, and because they do
not include explicitly the predator dynamics, the only bifurcation
occurring in those models is the flip bifurcation. Building on both modelling approaches, Myers et al.\cite{myers} have demonstrated that an unstable (decaying) period-4 oscillation can be excited by stochastic driving. However, it remains unclear over how long the oscillations can be maintained in that study. Furthermore, the frequently observed sequence of population numbers described in section 2 (and stably seen in our simulations, Fig. \ref{fig:timeseries}) only occurs episodically in the time series shown due to the stochastic nature of the approach.

In principle, the coupling between the four salmon lines can also
occur through the food of the sockeye fry. However, there is no
empirical evidence that there is a negative effect of a strong sockeye
year on the daphnia population in the following year that limits the
growth of the sockeye fry in that year\cite{levy}. 

The two conditions for a strong resonance of period 4 fit together
with the empirical observation that it occurs only in large
oligotrophic lakes, such as those of the Fraser River basin. The
smaller and ultra-oligotrophic lakes in the outer coast regions of British
Columbia do not show these oscillations. Their nutrient content is
neither large enough to raise large fry numbers, nor to allow for
salmon predators to become strong. 

On the other hand, cyclic dominance cannot be expected in
nutrient-rich lakes, because the spawning adults would not be an
important nutrient input, and because the coupling to the predator
would not be strong enough in a situation where there were more
predator species and more prey species for these predators.

Since there are large fluctuations in the proportion of fry that
survive to return to their nursery lake, we included noise in the
matching condition for the sockeyes (first Eq. 3 in \textit{Materials and Methods}) in order to determine with how much noise superimposed on the deterministic dynamics the
period-4 oscillation can persist. Fig. 2b shows a data series
generated with 50 percent noise in ocean survivability superimposed to
the deterministic dynamics. The oscillation is still clearly visible,
although the system has a phase shift every 300 years on average. With
less noise, the phase shift occurs less often, and with 100 percent
noise, the cyclic dominance vanishes in our simulations. When a large
perturbation acts only for a limited time, the oscillation quickly recovers
afterwards. In fact, the recovery following the blocking of the Fraser
River migration routes early in the last century, can be seen in the
non-dominant brood lines of the Shuswap stock in Fig. 1 and in Fig. A1 in Appendix A. Unfortunately, the year 2009 has seen
another large perturbation, with most Fraser sockeye expected to return that year not surviving in
the ocean, so that the expected strong escapement did not occur.

The age composition of sockeye stocks was also found to strongly influence the potential of the system to show a strong resonance. The fraction of adult sockeyes that returns at the age of 5 years instead of 4 years, $\epsilon$, was set to 0.1 in the simulations and we could show that the resonance appears for $0<\epsilon\leq 0.3$, but for very small values of $\epsilon$ ($<0.02$) the non-dominant lines disappear completely with increasing $K_0$. The resonance condition was best met at $\epsilon\approx 0.2$, where the resonance occurred nearly directly after the Neimark Sacker bifurcation. The parameter $\epsilon$ is also measured in the real populations\cite{healey}. For the sockeye salmon populations of the Fraser River it is approximately 0.08, while for the less productive outer coast lakes of British Columbia (where cyclic dominance is not observed) it is between 0.56 and 0.76. 

Some of the sockeye populations of the Bristol Bay area (Alaska), most notably the Kvichak River stocks, also exhibit strong oscillations\cite{fair,rogers-schindler} despite a broader distribution of the age at spawning than in the Fraser River stocks\cite{west1,west2}. However, the oscillations are not as regular, with maxima of the populations occurring every fourth or fifth year. This is consistent with our model, as it indicates a quasiperiodic oscillation rather than a fixed oscillation period associated with a strong resonance.

Our results do not rule out additional mechanisms such as depensatory
fishing\cite{walters-staley} or genetic effects \cite{walters-woodey}, which
would reduce the population sizes of weak lines to values smaller than those
resulting from our model. However, these additional assumptions are not
needed to explain the occurrence of cyclic dominance in the first place.

\section{Materials and Methods}
 \subsection{Model equations}
The dynamics of the biomass of sockeye fry $s_n(t)$, of their
predator $p_n(t)$, and of their zooplankton food $z_n(t)$ in year number $n$ during the growth season from spring ($t=0$) to fall ($t=T$) are given
by the following equations:
\begin{eqnarray}
\frac{d}{dt} s_n (t) &=& \lambda \cdot  a_{sz}\frac{z_n(t)\cdot s_n(t)}{1+c_s\cdot s_n(t)+z_n(t)} \nonumber \\&&- a_{ps}\frac{s_n(t)\cdot p_n(t)}{1+c_p\cdot p_n(t)+s_n(t)} - d_s\cdot s_n(t) \nonumber\\
\frac{d}{dt} z_n (t) &=& z_n(t)\cdot\left(1-\frac{z_n(t)}{K_n}\right)\\&&-a_{sz}\frac{z_n(t)\cdot s_n(t)}{1+c_s\cdot s_n(t)+z_n(t)}  \nonumber\\
\frac{d}{dt} p_n (t) &=& \lambda\cdot a_{ps}\frac{s_n(t)\cdot p_n(t)}{1+c_p\cdot p_n(t)+s_n(t)} - d_p\cdot p_n(t) \nonumber
\end{eqnarray}
The feeding terms include saturation at high prey densities, and a
predator interference term in the denominator (Beddington functional
response\cite{beddington,skalski}). Similar predator interference terms are also used by other
modellers\cite{walters}. 

The matching conditions used to determine the biomasses of the species at the beginning of the next season from their values at the end of the previous season(s) are given by
\begin{eqnarray}
s_{n+1}(0)&=&\gamma((1-\epsilon) s_{n-3}(T)+\epsilon s_{n-4}(T))\nonumber\\
z_{n+1}(0)&=&K_{n+1}\\
p_{n+1}(0)&=&p_n(T)\,,\nonumber
\end{eqnarray}
with $\epsilon$ the proportion of surviving sockeye that return
to their native lakes at the age of 5 to spawn and die. The carrying
capacity of the zooplankton in the next season, $K_{n+1}$, shows a
saturating dependence on the nutrient input due to the number of
spawning adults (and thereby on $ s_{n+1}(0)$),
\begin{equation}
 K_{n+1}=K_0+\left(\delta\frac{s_{n+1}(0)}{\delta_0+s_{n+1}(0)}\right)\, .
\end{equation}

\subsection{Parameters used for the computer simulations}
$\lambda=0.85$ denotes the assimilation efficiency of ingested prey biomass of carnivores. $a_{sf}=10$ and $a_{ps}=1.6$ are the maximal per unit biomass ingestion rates of salmon and predators, respectively. $d_s=1$ and $d_p=0.16$ represent biomass loss due to respiration and mortality. The metabolic rates of the predators are smaller than the respective rates of the salmon to account for the larger body size of the predators \cite{brown}; we assume the predators to be approximately 1500 times heavier than the sockeye fry. The interference parameters are set to $c_s=1.0$ and $c_p=0.2$. The parameter $\gamma$ in Eq. 3 summarises survival from smolt to adult fish (including ocean survivability and loss to fisheries), spawning success, and egg to fry survival. The various factors are estimated from empirical data \cite{hume1996,walters-staley,pauley} and yield $\gamma=0.4$. The 3 parameters determining the carrying capacity $K_n$ of the zooplankton in year $n$ are set to $K_0=15$, $\delta=10$, and $\delta_0=5$.

\section*{Acknowledgements}
This work was supported by the German Research Foundation (DFG) under
contract number Dr300/7 and Br2315/9-1.  Kim Hyatt, Jeremy Hume, and
Carl Walters commented on an earlier version of this
manuscript. Empirical data were kindly provided by Tracy Cone
(escapement data) and by Kim Hyatt (age composition of the sockeye
stocks). This collaboration started at the Complex Systems Workshop in
Fairbanks, Alaska, 2007. The 2008 Cultus Lake Modelling Workshop,
which had an important impact on this project, was hosted by Fisheries
and Oceans Canada.

\onecolumn
\makeatletter \renewcommand{\thefigure}{A\@arabic\c@figure}
\setcounter{figure}{0}
\appendix{\section{Supporting empirical data}\paragraph*{Escapement data for sockeye stocks showing cyclic dominance}
Many spawning stocks of sockeye salmon in the upper Fraser River basin are not stationary but have been dramatically increasing in size over the last decades (e.g. Fig. S1 a and b), following a massive disturbance of the stocks at the beginning of the 20th century. Nevertheless, the oscillatory pattern has clearly emerged again. 

\begin{figure}[h!t!b!p]
 \begin{center}
  \includegraphics[scale=1, angle=0]{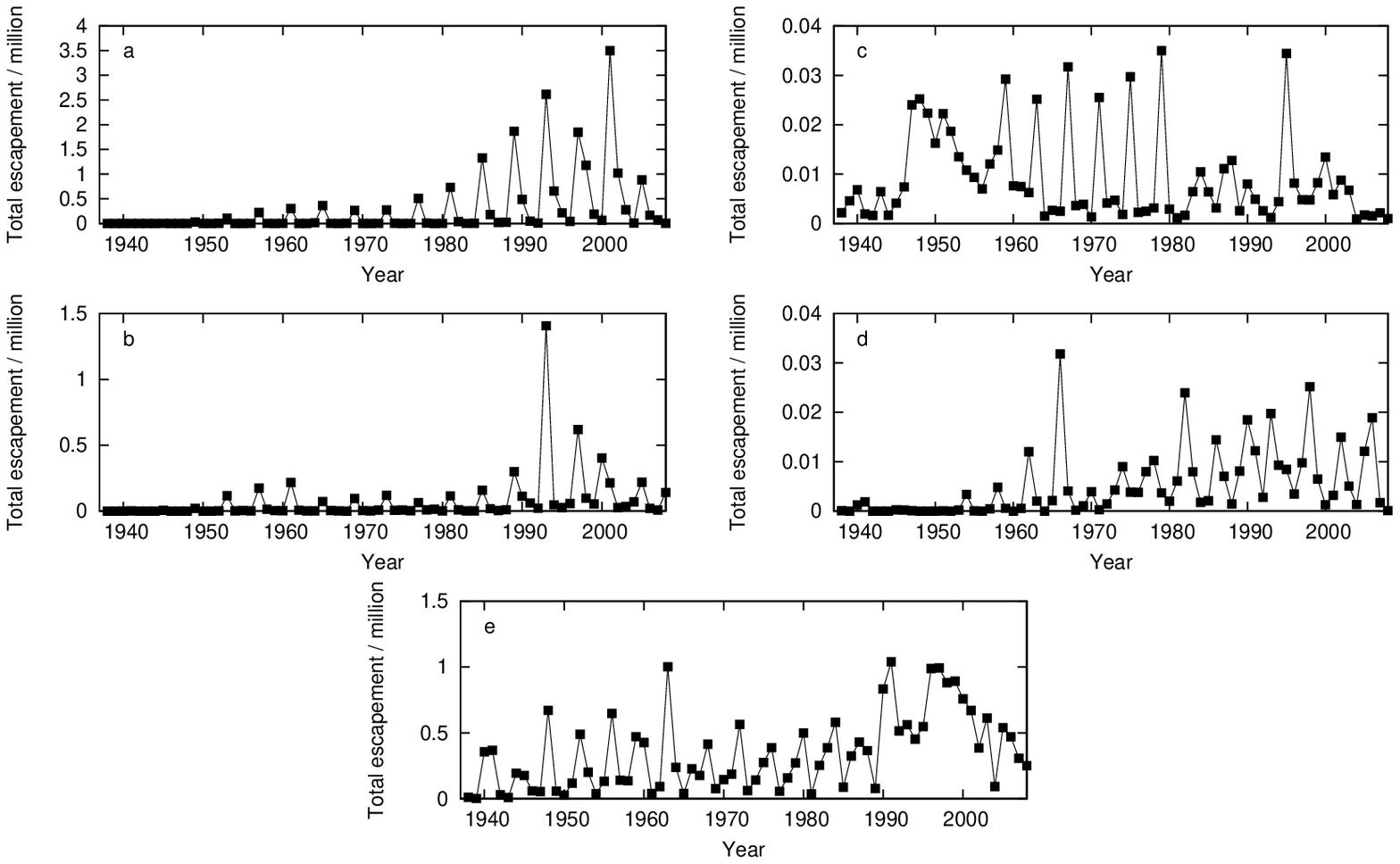}
  \end{center}
\caption{Escapement data of for sockeye spawing stocks in the Fraser River basin. a: Quesnel Lake, b: Stuart Lake, c: Bowron Lake, d: Seton Lake, e: Chilko Lake.}
\end{figure}
}

\end{document}